# Structural and Electrical Properties of Grafted Si/GaAsSb Heterojunction

Haris Naeem Abbasi[1], Seunghyun Lee[2], Hyemin Jung[2], Nathan Gajowski[2], Yi Lu[1], Linus Wang[1], Donghyeok Kim[1], Jie Zhou[1], Jiarui Gong[1], Chris Chae[2], Jinwoo Hwang[2], Manisha Muduli[2], Subramanya Nookala[1], Zhenqiang Ma[1, *], and Sanjay Krishna[2, *]

[1]*Department of Electrical and Computer Engineering, University of Wisconsin-Madison, Madison, Wisconsin, 53706, USA*

[2]*Department of Electrical and Computer Engineering, The Ohio State University, Columbus, Ohio 43210, USA*

[*]Author to whom correspondence should be addressed. mazq@engr.wisc.edu and krishna.53@osu.edu

The short-wave infrared (SWIR) wavelength, especially 1.55 µm, has attracted significant attention in various areas such as high-speed optical communication and LiDAR systems. Avalanche photodiodes (APDs) are a critical component as a receiver in these systems due to their internal gain which enhances the system performance. Silicon-based APDs are promising since they are CMOS compatible, but they are limited in detecting 1.55 µm light detection. This study proposes a *p*-type Si on *n*-type $GaAs_{0.51}Sb_{0.49}$ (GaAsSb) lattice matched to InP substrates heterojunction formed using a grafting technique for future GaAsSb/Si APD technology. A $p^+$Si nanomembrane is transferred onto the GaAsSb/AlInAs/InP substrate, with an ultrathin ALD-$Al_2O_3$ oxide at the interface, which behaves as both double-side passivation and quantum tunneling layers. The devices exhibit excellent surface morphology and interface quality, confirmed by atomic force microscope (AFM) and transmission electron microscope (TEM). Also, the current-voltage (I-V) of the $p^+$Si/$n^-$GaAsSb heterojunction shows ideal rectifying characteristics with an ideality factor of 1.15. The I-V tests across multiple devices confirm high consistency and yield. Furthermore, the X-ray photoelectron spectroscopy (XPS) measurement reveals that GaAsSb and Si are found to have type-II band alignment with a conduction band offset of 50 meV which is favorable for the high-bandwidth APD application. The demonstration of the GaAsSb/Si heterojunction highlights the potential to advance current SWIR PD technologies.

Photodetectors (PDs) operating in the short-wave infrared (SWIR) spectrum have attracted considerable attention across various fields, including remote sensing, telecommunications, and laser guidance systems [1-3]. Among these applications, the 1.55 µm wavelength has emerged as a center point for photodetector development due to its compatibility with mature fiber optic technologies, making it suitable for data centers and optical communication systems [3-5]. These systems rely on lasers for transmitting data and detectors for receiving it. However, over long distances, the input signal may weaken, necessitating signal amplification. In such cases, avalanche photodiodes (APDs) are an excellent choice for the receiver part due to their internal gain, which enhances the signal-to-noise ratio and provides a higher gain-bandwidth product[6-8].

Silicon (Si) stands out as the prime material for the avalanche multiplication layer. Si-based APDs exhibit high gain and low excess noise, making them an attractive choice for applications in optical communication, light detection, and ranging (LIDAR) and optical sensors [9-11]. However, Si has a limitation on detecting the 1.55 µm wavelength light due to its bandgap nature[12]. To overcome this limitation, extensive research has explored the heterogeneous integration of III-V absorber materials with Si multiplier. Previous research has investigated Ge/Si APDs using epitaxial integration, which have demonstrated a high gain-bandwidth product (GBP) >860 GHz at a wavelength of 1310 nm [9]. However, a significant lattice mismatch between Si and Ge leads to the accumulation of dislocations during the material growth, resulting in high dark current [10, 13, 14]. Additionally, the integration of InGaAs/Si APDs using wafer bonding techniques has shown promise in combining Si's low-noise and high-gain performance with the excellent light-absorbing properties of the InGaAs absorber [15]. Although the wafer bonding method has shown potential, its high cost and interface defects along with the large conduction band offset (~ 0.33 eV)[16] between InGaAs and Si necessitate a better approach.

Alternatively, an approach involves using $GaAs_{0.51}Sb_{0.49}$ (hereafter GaAsSb), which is lattice matched to InP substrates and can serve as a substitute for the InGaAs absorber [17-19]. GaAsSb shares similar bandgap and absorption coefficient characteristics with InGaAs, making it a viable option[6]. Moreover, GaAsSb presents a theoretically well-matched conduction band offset with Si, enhancing carrier transport. Therefore, GaAsSb can address a crucial challenge in III-V/Si APD integration [18]. To achieve this integration successfully, we employed the grafting technique to transfer Si on GaAsSb[20]. This pioneering technique enables the integration of materials with lattice mismatches by utilizing an ultrathin interfacial oxide layer that chemically bonds two dissimilar semiconductors while avoiding their direct lattice contact, which would otherwise introduce elevated interface traps. This ultrathin oxide layer serves two purposes, as a passivation layer to suppress interfacial traps and as a quantum tunneling layer to facilitate charge transport. The grafting technique has successfully facilitated heterojunctions between various lattice-mismatched semiconductors, including Ge/Si, Si/GaAs, Si/GaN, GaAs/GaN, $Si/Ga_2O_3$, $GaAs/Ga_2O_3$, and GaAs/Diamond etc. [20-25].

In this work, we demonstrate a *p*-type Si/*n*-type GaAsSb PN heterojunction using the grafting technique. We conducted a comprehensive characterization of the transferred devices using techniques including atomic force microscopy (AFM), transmission electron microscopy (TEM), and current-voltage (IV) measurements to assess the structural and electrical properties of the integrated Si/GaAsSb PN devices. Furthermore, the band alignment of GaAsSb and Si is experimentally measured by X-ray photoelectron spectroscopy (XPS). This demonstration marks a significant step toward the development of GaAsSb/Si APD technology in the future.

In Figure 1(a), the schematic structure of the n-GaAsSb absorber segment is presented. The material growth of the device stack was conducted on a 2-inch epi-ready, n-type, on-axis InP substrate using a RIBER compact 21DZ solid-state molecular beam epitaxy (MBE) system. The preparation of a growth-ready surface on the InP substrate is quite tricky, primarily due to the presence of a thin native oxide layer. Typically, the existence of native oxides in other material systems, such as GaAs and GaSb, can be ascertained using the in-situ reflection high-energy electron diffraction (RHEED) technique within an MBE system. However, the observation of oxide-related RHEED patterns becomes exceptionally challenging when growing on an InP substrate. Furthermore, in the oxide removal process, an arsenic (As) soak is commonly employed, as opposed to a phosphorus (P) soak, due to the limited availability of favorable P sources within MBE equipment, thereby hindering the clear observation of RHEED patterns. As a result, the removal of the native oxide from the InP substrate was achieved by detecting the RHEED transition from a 2×4 (As-rich) to a 4×2 (In-rich) pattern, signifying the complete elimination of the oxide layer. Subsequently, a 250 nm thick unintentionally doped (UID) $In_{0.52}Al_{0.48}As$ (InAlAs) buffer layer was grown to initiate the overall device structure while maintaining a smooth surface morphology. Following this, a 100 nm thick heavily doped (~$10^{19}$ $cm^{-3}$) $n^{++}$-GaAsSb layer was deposited as the n-type contact layer, succeeded by a 300 nm thick lightly doped (~$10^{17}$ $cm^{-3}$) $n^{-}$-GaAsSb layer for the formation of a pn junction with a $p$-Si NM.

A 6-inch Soitec(R) p-type silicon-on-insulator (SOI) substrate with a 205 nm Si device layer and a 400 nm BOX layer was thermally oxidized at 1050 °C for 12 minutes, creating a 36 nm thick screen oxide. Boron was then implanted at room temperature ($3 \times 10^{15}$ $cm^{-2}$, 15 keV, 7° tilt) and activated at 950 °C for 40 minutes, reducing the Si device layer to 180-185 nm with boron concentrations between $9.7 \times 10^{19}$ $cm^{-3}$ (top) and $9.5 \times 10^{19}$ $cm^{-3}$ (bottom), guided by Silvaco Athena simulations. The process preserved the monocrystalline nature of the $p^+$ Si layer, with the screen oxide thickening to ~52 nm [23].

To fabricate the Si/GaAsSb diode, the $p^+$ silicon wafer was cut into 4×4 $mm^2$ pieces using a Disco Dicing Saw. To release the nanomembranes (NMs) of $p^+$Si from the epitaxial wafer, these cut pieces underwent a standard semiconductor cleaning process. This process involved rinsing the sample with acetone, isopropyl alcohol (IPA), and deionized (DI) water for 10 minutes each under ultra-sonication, followed by drying with nitrogen ($N_2$) gas. Subsequently, a layer of photoresist (PR) 1813 was applied to the samples, and 9×9 $\mu m^2$ holes were patterned using a mask and UV-photolithography. The reactive ion etcher (RIE) (Plasma-Therm) was then used to etch through the patterns down to the sacrificial $SiO_2$ layer, using RF power of 100W with $SF_6$ gas. Afterward, the PR was removed using acetone and IPA. Finally, the samples were soaked in a 49% HF solution for 3 hours to undercut the $SiO_2$ sacrificial layer, with acetone rinses to expedite the process. After completing the undercut, any remaining HF residues were removed with DI water, and the samples were dried using $N_2$. The $p^+$Si NM was then lifted-off using a polydimethylsiloxane (PDMS) stamp. Then, the NM was transferred to the GaAsSb substrate where the substrate surface was cleaned using acetone, IPA, and DI water, followed by immersion into the diluted HF solution, and then a 0.5 nm $Al_2O_3$ layer was deposited using atomic layer deposition (Savannah S200 ALD, USA). Finally, the $p^+$Si nanomembranes on the substrate underwent rapid thermal annealing (RTA) at an optimized temperature of 350°C for 5 minutes to turn the weak bonding between the Si NM and the GaAsSb substrate into robust chemical bonding. This process did not result in any degradation or noticeable changes in the quality of the Si NM as discussed in the following section. The complete NM transfer process is shown in our previous work [25].

The bottom and top metal contacts were deposited using an Angstrom Engineering Nexdep Physical Vapor Deposition Platform (e-beam) tool. Initially, for the cathode mesa, the sample underwent patterning through photolithography. Subsequently, selective etching was carried out using an RIE tool from Plasma-Therm. This involved using RF power of 100 W with $SF_6$ gas for Si and RF power of 60 W with a mixture of $BCl_3$ (10 SCCM) and Ar (5 SCCM) at a working pressure of 15mT for GaAsSb. This process etched through the $p^+$Si NM and $n^-$GaAsSb down to the $n^+$GaAsSb base region. Next, the cathode pattern was defined using photolithography, and a stack of metals (Pd/Ge/Au: 30/40/100 nm) was deposited at a constant rate of 0.5 Å per second using the e-beam technique with a power of 500 W. The base pressure during this deposition was maintained at $5\times10^{-7}$ torr. Moving on to the anode metal deposition, the anode pattern was defined through lithography, and a metal stack of Ni/Au with a thickness of 10/100 nm and low contact resistivity [26] was deposited using the same e-beam evaporator. After the metal deposition for both the cathode and anode, individual diodes were isolated by lithography patterning and RIE etching, using $SF_6$ gas and an RF power of 100 W for 2 minutes. Finally, the device was passivated using an 8 nm layer of $Al_2O_3$ deposited via ALD. The entire fabrication process flow for the device is shown in Figure 1 (a) to (f).

The band alignment of the Si/GaAsSb heterostructure was characterized using a Thermo Scientific K Alpha XPS equipped with an Al Kα X-ray source (hν = 1486.6 eV). Experimental parameters included a 50 eV pass energy, 400 μm spot size, dwell times of 0.5 s, and a step size of 0.05 eV. The binding energy of the C 1s peak (284.8 eV) served as the standard reference for core-level calibration. Three distinct samples were prepared: pristine Si, pristine GaAsSb, and the grafted Si/GaAsSb heterostructure for the band alignment study. However, due to the limited sampling depth of XPS, approximately 10 nm beneath the sample surface, an additional top-down dry etching step was necessary to unveil interface information for the grafted Si NM. This process was executed using a PlasmaTherm 790 RIE etcher, ensuring precise control over the etching rate. Consequently, the Si NM was thinned from its original thickness of 180 nm down to <10 nm while preserving its single crystallinity. Moreover, the Si sample underwent a 5-minute immersion in a 49% HF solution, on the contrary, the GaAsSb samples were dipped into diluted HF. Afterward, all samples were promptly transferred to a nitrogen-filled and sealed zip bag before loading into the XPS chamber, aiming to minimize native oxide formation.

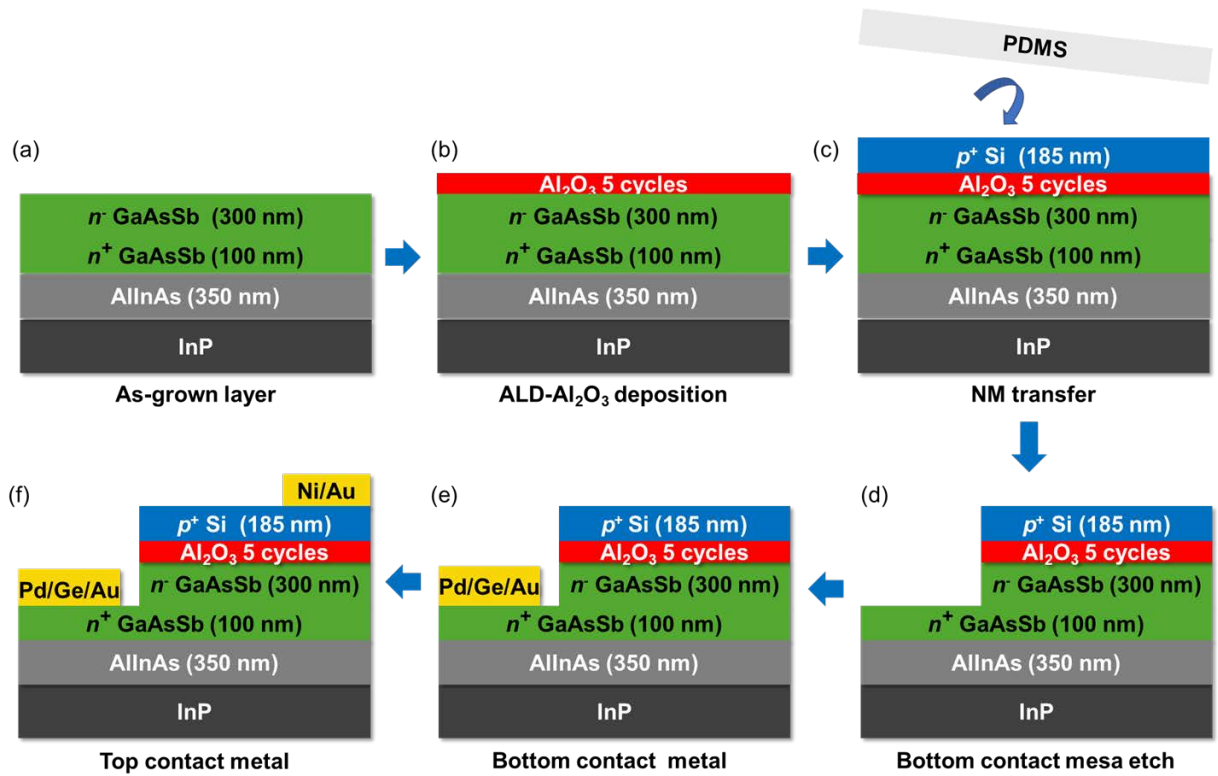

Fig 1. Process flow of the fabrication. (a) Substrate dicing and cleaning. (b) 0.5 nm $Al_2O_3$ deposition using atomic layer deposition. (c) $p^+$Si nanomembranes transfer using the polydimethylsiloxane (PDMS) stamp. (d) Cathode mesa etching using dry etching technique. (e) Cathode metal deposition using ebeam evaporator. (f) Anode metal deposition using ebeam deposition.

Figure 2 (a) presents an optical microscopy image depicting the Si NM transferred onto a GaAsSb substrate. The image highlights six rectangular regions each measuring 9µm × 9µm, which were the areas designated for selective etching of the $SiO_2$ sacrificial layer. The resultant surface is smooth without any apparent bubbles. For a more detailed examination, an atomic force microscopy (AFM) image was captured, as shown in Figure 2 (b). The measured surface roughness is 0.433 nm over a 3 µm scale, indicating a transfer of high-quality Si NM onto GaAsSb. The Si NM thickness, estimated to be approximately 185 nm by AFM step height measurement, is also shown in Figure 2 (c), aligning with the target thickness. Additionally, Figure 2 (d) shows a scanning electron microscopy (SEM) image of the Si/GaAsSb devices after device fabrication, clearly observing the top contact on $p^+$Si and bottom contact on $n$-GaAsSb and they are quite uniform.

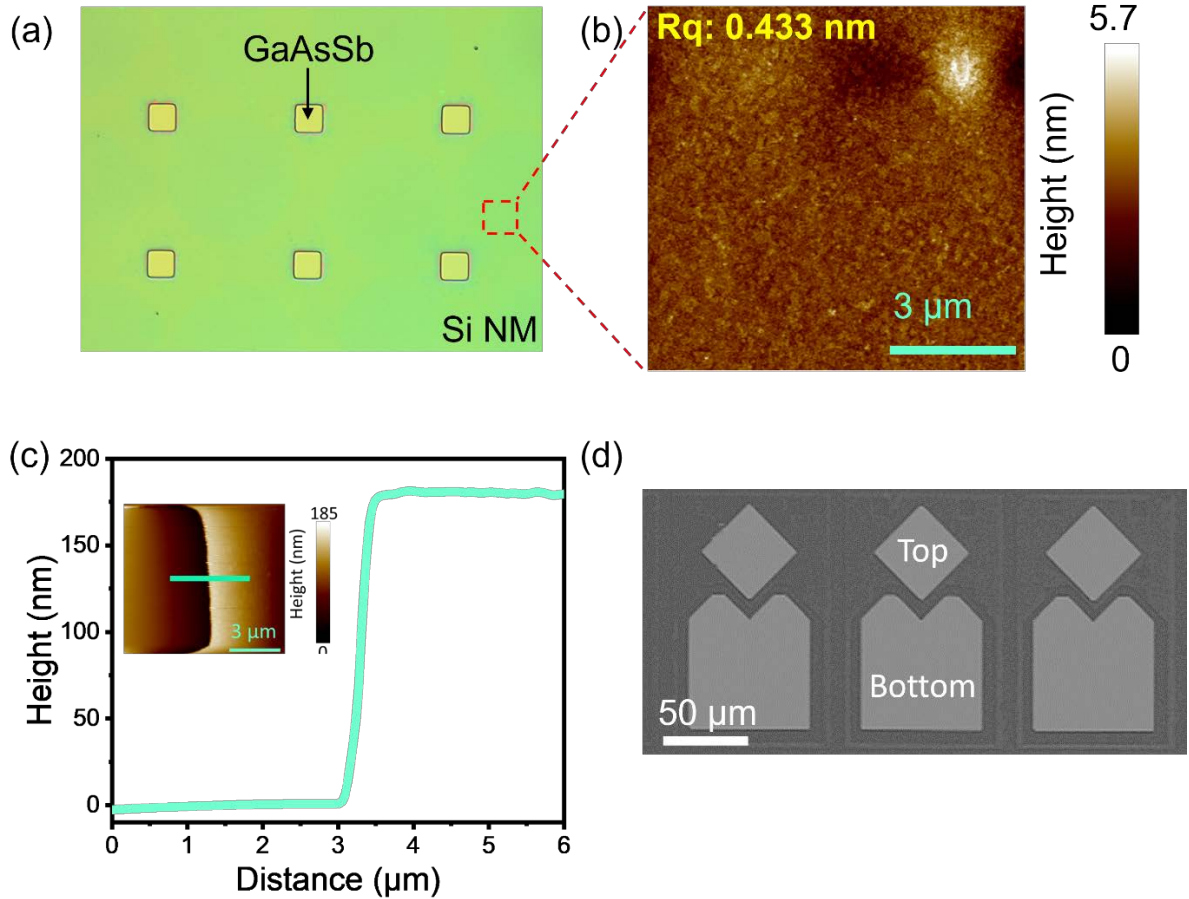

Fig 2. Morphology analysis of nanomembranes (NM) and device. (a) Optical microscopy image of NM showing the holes and clean surface. (b) Atomic force microscopic images at 5×5 µm² area, with an inset RMS (Rq) roughness value of 0.433 nm. (c) Atomic force microscopic step height calculation at 10×10 µm² area with a NM thickness of ~185 nm. (d) Scanning electron microscopic image of the three $p^+$Si/$n^-$GaAsSb devices, with a scale bar of 50 µm.

In typical heterogeneously integrated devices, the focus is often not on their function as an active interlayer. Rather, the precise placement on a silicon-based integrated circuit is crucial, without the responsibility for carrier transport across the bonded heterointerface. However, the Si/GaAsSb PN junction device investigated in this research necessitates carrier transport between the layers to enhance carrier collection efficiency at both top and bottom contacts. Additionally, at the operational bias, it is imperative that the device encounters no barriers that could adversely affect its bandwidth. Consequently, a key area of investigation is the structural and electrical interface quality between Si and GaAsSb.

Structural analysis of the Si/GaAsSb interface was conducted using high-resolution transmission electron microscopy (HRTEM). Figure 3 (a) reveals a cross-sectional view of the device comprising four distinct layers: InP, AlInAs, GaAsSb, and Si. An in-depth look at the InAlAs/GaAsSb interface is provided in Figure 3 (b), showing the two ternaries, InAlAs and GaAsSb, are lattice-matched to InP, indicative of excellent material growth. TEM images in Figure 3 (c) confirm a sharp interface and crystalline perfection, indicating the successful transfer and robust chemical bonding of Si NM to GaAsSb, without visible air voids, or defects like misfit dislocations. An expanded image delineates the presence of some amorphous layers apart from a thin $Al_2O_3$ oxide layer. The interface layer's total thickness is around 5 nm. This widened ALD-$Al_2O_3$ interlayer deviated from the as-deposited thickness (~0.5 nm) possibly due to the redistribution of oxygen atoms forming chemical bonding with the surrounding materials during the thermal annealing process. Optimizing this interface layer's thickness is vital to enhance carrier collection, as a thinner oxide layer increases the probability of carrier tunneling, leading to a more efficient device. Nevertheless, the carriers can be transported via a quantum tunneling mechanism.

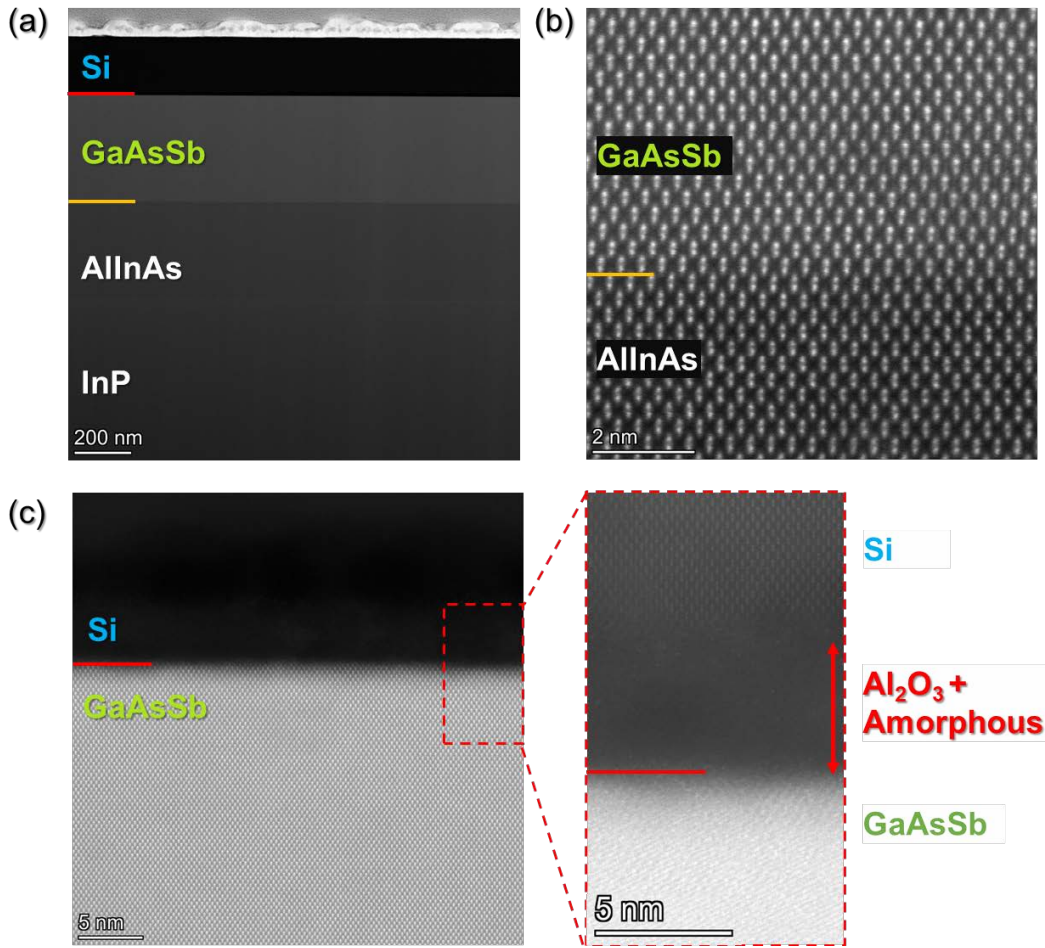

Fig 3. (a) The cross-sectional TEM image of Si NM transferred on GaAsSb segment. (b) TEM image of the interface between GaAsSb and InAlAs layers. (c) TEM image of the interface between Si and GaAsSb with an enlarged image on the right side.

Electrical characterization of the Si/GaAsSb interface was conducted through current-voltage (IV) measurements in the dark condition, ranging from -2 to 2 V at room temperature, as shown in Figure 4. The device displays rectifying characteristics with a low turn-on voltage of 1.4 V obtained from linear extrapolation [27] (Figure 4 (a)) and a high rectification ratio of $1.41 \times 10^2$ (Figure 4 (b)) at ±2 V, indicating a high-quality interface. A heterojunction diode's performance can be partly identified by its ideality factor, a metric reflecting the deviation from an ideal diode [28]. This factor is instrumental in examining the carrier recombination mechanism and the junction interface's quality. The ideality factor for the Si/GaAsSb heterojunction device, derived from the semilog IV curve's slope under low forward bias as shown in Figure 4 (b), was found to be 1.15. Despite being slightly above the ideal, this minor deviation is attributable to the interface oxide layers introduced during the thermal annealing process, as discussed in Figure 3. The near-ideal ideality factor underlines the excellence of the NM transfer process. Figure 4 (c) compiles the IV curves from various Si/GaAsSb devices, with an average ideality factor of 1.23 calculated from ten randomly selected devices, demonstrating consistency and high uniformity across the sample.

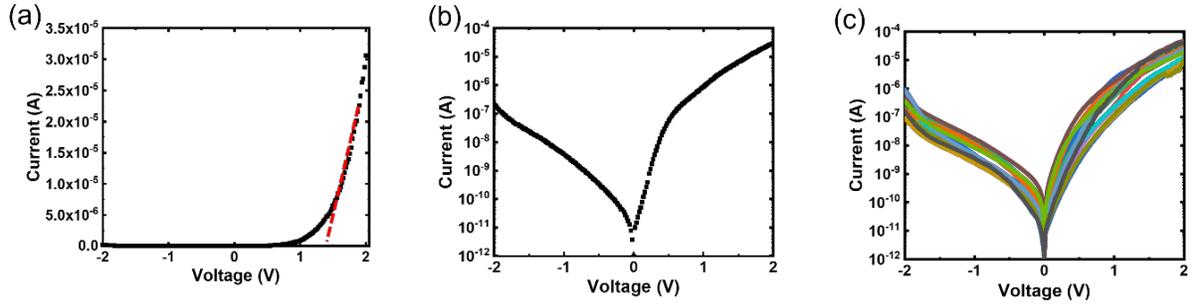

Fig. 4. Current-Voltage measurements of the device. (a) linear scale and (b) logarithmic scale I-V measurements with an inset image of ideality factor (IF) and On/Off ratio. (c) Multiple devices IV measurements in log scale showing uniformity.

To ascertain the band alignment of the Si/GaAsSb heterostructure, Kraut et. al's [29, 30] approach was applied. This XPS analysis validates theoretical predictions of band alignment and assesses interface quality. A high density of interface traps ($D_{it}$), common in directly wafer-bonded interfaces, can cause Fermi-level pinning, leading to discrepancies between experimental band alignment and theoretical models [25, 31]. To enable accurate XPS characterization, the Si/GaAsSb heterojunction sample was processed to reduce the thickness of the Si nanomembrane. The valence band offset, VBO ($\Delta E_v$) of Si and GaAsSb could be calculated from the following formula [29]:

$$\Delta E_v = \left(E_{Si\ 2p}^{Si} - E_{VBM}^{Si}\right) - \left(E_{Sb\ 3d}^{GaAsSb} - E_{VBM}^{GaAsSb}\right) + \left(E_{Sb\ 3d}^{GaAsSb} - E_{Si\ 2p}^{Si}\right)_{interface} \quad (1)$$

$E_{Si\ 2p}^{Si}$ and $E_{Sb\ 3d}^{GaAsSb}$ represent the Si 2p and Sb 3d binding energy position of pristine Si and GaAsSb film, while $E_{VBM}$ denotes the valence band maximum (VBM) position of respective samples. In Figure 5 (a) and (c), $E_{Si\ 2p}^{Si}$ and $E_{Sb\ 3d}^{GaAsSb}$ exhibit core level peak positions at 99.76 and 537.43 eV, respectively. Figures 5 (b) and (d) illustrate the valence band spectrum, where VBM is estimated by linearly extrapolating the leading edge to the baseline of the respective valence band photoelectron spectrum. The VBM values of pristine Si and GaAsSb are measured to be 0.89 and 0.43 eV. Figure 5 (e) and (f) display the Si 2p and Sb 3d core levels with peak values of 99.87 and 537.57 eV, respectively, from the grafted Si/GaAsSb heterostructure. By incorporating all these experimental values into equation (1), the VBO for the Si/GaAsSb heterostructure is determined to be -0.43 eV.

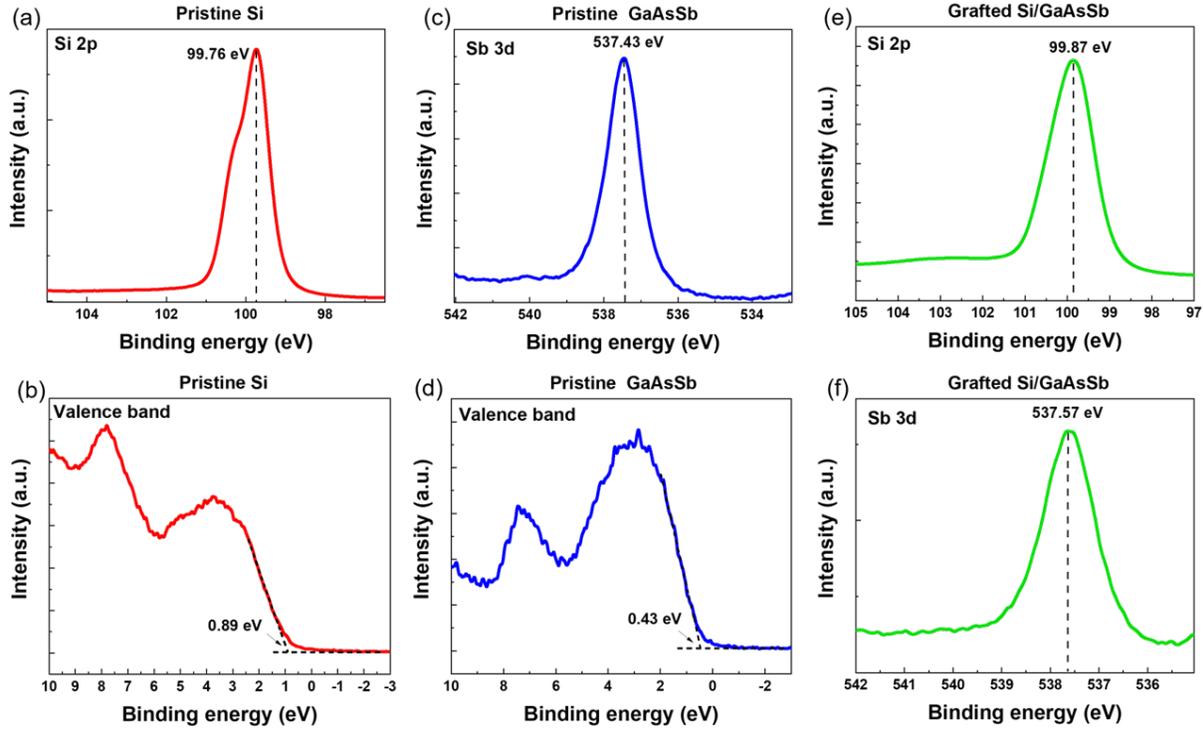

Fig. 5. XPS core level spectra of (a) Si 2p and (b) valence band spectra of the pristine Si; core level spectra of (c) Sb 3d and (d) valence band spectra of the pristine GaAsSb; core level spectra of (e) Si 2p and (f) Sb 3d of the grafted Si/GaAsSb interface.

The conduction band offset, CBO ($\Delta E_c$) for Si/GaAsSb heterostructure is determined by substituting the VBO and energy bandgap ($E_g$) values of Si and GaAsSb in equation 2,

$$\Delta E_c = E_g^{Si} - E_g^{GaAsSb} - \Delta E_v \qquad (2)$$

$E_g^{Si}$ and $E_g^{GaAsSb}$ values are 1.12 and 0.74 eV[32], respectively. Hence, the CBO is estimated to be -0.05 eV with an error of $\pm 0.05$ eV, forming a type-II heterostructure, which is shown in Figure 6.

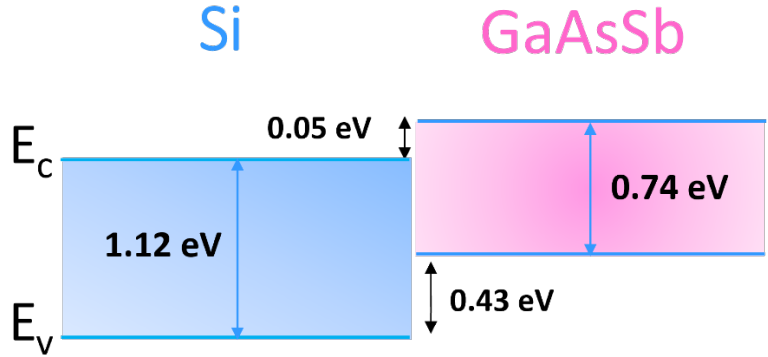

Fig. 6. The band alignment diagram of the grafted Si/GaAsSb heterojunction.

The theoretical and experimental values for the VBO and CBO reveal a close agreement, indicating a strong correlation between predicted and measured results. The theoretical $\Delta E_c$ is 0.02 eV, while the experimental $\Delta E_c$ is slightly higher at 0.05 eV. Similarly, the theoretical $\Delta E_v$ is 0.36 eV, compared to the experimental value of 0.43 eV. The theoretical values were obtained using electron affinities of Si and GaAsSb as 4.05 eV and 4.07 eV, respectively[33, 34]. The close alignment between theoretical and experimental values underscores the effectiveness of the grafting technique, suggesting that the heterojunctions were formed with minimal interface defects.

Furthermore, Type-II band alignment can offer advantages in APD development. A key factor in the design of APDs is the uninterrupted flow of carriers. Typically, a grading layer is used to bridge the absorber and the multiplication layer to circumvent potential carrier blockage caused by a large conduction band offset. Nevertheless, when employing the type-II alignment found in GaAsSb/Si combinations, the grading layer is no longer required, simplifying the APD design process and its structure. Additionally, the absence of a barrier promotes smoother carrier transport, potentially broadening the APD's bandwidth and indicating its suitability for high-bandwidth applications.

In summary, this study investigates the integration of GaAsSb with Si. The fabrication method includes transferring a $p^+$Si NM onto an $n^-$GaAsSb/AlInAs/InP wafer, yielding a device of exceptional quality. The devices demonstrate high surface and interface qualities, as verified by AFM and TEM. IV assessments reveal linear characteristics, characterized by an ideality factor of 1.15 and an on/off ratio of $1.41 \times 10^2$. Tests on device uniformity show reliable performance across multiple devices. The oxide's thickness at the interface is crucial for carrier tunneling, enhancing electrical performance, quantum efficiency, and bandwidth. However, any unintended amorphous layer beyond what is needed for double-side passivation can create non-uniform interfaces, reducing tunneling efficiency and device performance. This irregularity can cause localized high electric fields in APDs, leading to dielectric breakdown and misfit dislocations, resulting in premature device failure. Therefore, controlling the amorphous layer's thickness is essential for ensuring the high performance and reliability of NM-transferred devices, especially in sensitive and high-speed applications. Moreover, the experimentally observed type-II band alignment between GaAsSb and Si promotes improved carrier transport, beneficial for high-bandwidth SWIR APDs. Consequently, the GaAsSb/Si heterojunctions offer a solution to the current limitation of Si-based APDs, paving the way for progress in SWIR imaging and communication systems.

**ACKNOWLEDGMENTS**

Funded by the Intel® Semiconductor Education and Research Program for Ohio. The device fabrication was partially supported by Defense Advanced Research Projects (DARPA) under grant HR00112190107 and by Air Force Research Lab under grant FA8649-22-P0736.

**DATA AVAILABILITY**

The data that support the findings of this study are available from the corresponding author upon reasonable request.

**AUTHOR CONTRIBUTIONS**

H. N. A. and S.L. contributed equally to this paper.